\documentclass[prb,superscriptaddress,twocolumn,showpacs,byrevtex,endfloats]{revtex4}
\usepackage{graphicx}
\begin{document}
\preprint{}
\title {\bf {Optical functions and electronic structure of
CuInSe$_2$, CuGaSe$_2$, CuInS$_2$, and CuGaS$_2$}}
\author{M.~I.~Alonso}
\email{Isabel.Alonso@icmab.es}
\affiliation{Institut de Ci\`encia de Materials de Barcelona,
CSIC, Campus de la UAB,
08193 Bellaterra, Spain}
\author{K. Wakita}
\affiliation{Department of Physics and Electronics, College of
Engineering,
Osaka Prefecture University, Gakuen-cho 1-1, Sakai, Osaka 599-8531,
Japan}
\author{J.~Pascual}
\affiliation{Departament de F\'{\i}sica, UAB, and Institut de Ci\`encia
de Materials de Barcelona, CSIC,
08193 Bellaterra, Spain}
\author{M.~Garriga}
\affiliation{Institut de Ci\`encia de Materials de Barcelona,
CSIC, Campus de la UAB,
08193 Bellaterra, Spain}
\author{N. Yamamoto}
\affiliation{Department of Physics and Electronics, College of
Engineering,
Osaka Prefecture University, Gakuen-cho 1-1, Sakai, Osaka 599-8531,
Japan}
\date{\today}
\begin{abstract}

We report on the complex dielectric tensor components of four chalcopyrite
semiconductors in the optical energy range (1.4--5.2~eV, from 0.9~eV for
CuInSe$_2$) determined at room temperature by spectroscopic ellipsometry.
Our results were obtained on single crystals of CuInSe$_2$, CuGaSe$_2$,
CuInS$_2$, and CuGaS$_2$.  Values of refractive indices $n$, extinction
coefficients $k$ and normal-incidence reflectivity $R$ in the two different
polarizations are given and compared with earlier data where available.  We
analyze in detail the structures of the dielectric function observed in the
studied energy region. Critical-point parameters of electronic transitions
are obtained from fitting of numerically calculated second-derivative
spectra $d^2\varepsilon (\omega)/d\omega^2$. Experimental energies and
polarizations are discussed on the basis of published band structure
calculations.

\end{abstract}
\maketitle
\section{Introduction}

The studied ternary compounds Cu-III-VI$_2$ (III = Ga, In and VI = S, Se)
are direct gap semiconductors with tetragonal chalcopyrite (CH) crystal
structure.  This family of materials is relevant in many fields, including
nonlinear optics, optoelectronic, and photovoltaic devices.
\cite{ShWe75,OhPa98,BiEs97,ChSh97} Accurate knowledge of the optical
functions of these materials is very important for many of these
applications.  In spite of the considerable amount of research devoted to
these materials, this knowledge is still incomplete. In this paper, we
present careful ellipsometric measurements over the energy range 1.4 to 5.2
eV (from 0.9 eV for CuInSe$_2$) that provide values of the complex
dielectric functions $\varepsilon (\omega) =
\varepsilon_1 (\omega) + i \varepsilon_2 (\omega)$ both in the ordinary and
extraordinary polarizations. We discuss our results taking into
account previous related work.

Another concern of this work is the understanding of the electronic
structure of these compounds, especially focusing on the origin of the
interband transitions above the band gap. Energies and selection rules of
the transitions, both observed in this work and reported in the literature
for the different studied compounds, are discussed. As basis, we consider
the band structure calculations of Jaffe and Zunger \cite{JaZu83} using a
self-consistent approach within the density-functional formalism. We find
common trends in the spectra of the four compounds, in agreement with the
mentioned calculation.  \cite{JaZu83} Despite the large influence of Cu-$3d$
states on the electronic band structure, the main optical transitions are
shown to originate between hybridized bands. Thus, these spectra bear a
rather close relationship with those of binary zinc blende (ZB) compounds in
general.

The paper is organized as follows. After a short description of the
experiments in Sect. \ref{expt}, the results are presented in the next two
Sections. First, in Sect. \ref{dielectric}, we report and discuss the values
of the optical functions of the four compounds. In Sect. \ref{cps} we
analyze in detail the structures of the dielectric function observed in the
studied energy region. Then, in Sect. \ref{elestr} we relate the
critical-point energies to the electronic band structures of the compounds.
Finally, we summarize the most important results in Sect. \ref{summary}.

\section{Experimental details}\label{expt}

The samples used in this study were single crystals.  In the case of
CuInSe$_2$ we measured a platelet with (112) orientation grown by chemical
iodine-vapor transport (IT).  The other three samples consisted of (001)
oriented faces cut from ingots grown by the traveling-heater method (THM).
The THM process requires use of a solvent that may incorporate as impurity
in the resulting crystal.  In this case, use of In solvent yielded crystals
of CuGa$_{1-x}$In$_x$(S,Se)$_2$ with small In contents $x$, and
stoichiometric CuInS$_2$. We have paid special attention to the problem of
removing surface overlayers, which is of primary importance in
spectroellipsometric measurements. We used the accepted criteria of Aspnes
and Studna \cite{AsSt83} to determine the optically ``best'' surfaces to
obtain dielectric function values representative of bulk semiconductors. The
best results for the IT-CuInSe$_2$ sample were obtained after etching of the
as-grown surface in a solution of 5\% hydrofluoric acid in de-ionized water.
The THM crystals were sequentially polished with slurries of successively
finer alumina powder (down to 0.3 $\mu$m grid size) in de-ionized water on
suitable polishing cloths. Immediately before spectroscopic ellipsometry
(SE) measurements, samples were chemomechanically polished with an alkaline
colloidal silica suspension (Buehler's Mastermet), rinsed with de-ionized
water, and blown dry with $N_2$. Variations of this procedure, such as
rinsing with methanol instead of water, or further chemical etching of the
surface, either did not modify or led to worse spectra, showing both lower
$\langle \varepsilon \rangle$ values \cite{AsSt83} and broader spectral
features.

Measurements were done using a spectral ellipsometer with rotating
polarizer, at room temperature, and keeping the sample under dry $N_2$ flux
to delay surface contamination. Depending on the sample, more or less
degradation was observable after several hours. The spectral range of all
measurements was 1.4 to 5.2~eV and for CuInSe$_2$ we also measured the
bandgap region down to 0.9~eV, using a GaInAs photodiode as detector. In all
spectral measurements, the angle of incidence was $\varphi=65^\circ$ and the
analyzer azimuth $A=20^\circ$. The experimental energy step was generally
20~meV, but we used finer meshes of 2 to 10~meV for the sharper gap
features. In the THM crystals we acquired two spectra with the plane of
incidence either parallel or perpendicular to the in-plane optical axis,
characterized by x-ray diffraction. The orientation of the IT sample was
checked optically by $\beta$-scan measurements \cite{AlGa95} to determine
the needed projection of the direction of the $c$ axis on the sample
surface. We measured $\beta$-scans at two energies with pronounced
anisotropy (2.8 and 4 eV) and three analyzer settings (5, 10, and 30
degrees). The obtained results followed well the behavior of a uniaxial crystal
with its optic axis forming an angle $\alpha$ with the surface normal. The
fitted angle was $\alpha = 55 \pm 1 ^\circ$, in good agreement with a (112)
surface and optic axis along [001]. For this sample we took four spectral
measurements at Euler's angles $\beta = $ 0, 90, 180, and 270 degrees to
extract the tensor components. Both if the optic axis is on the sample
surface or not, there is no direct analytical expression relating the
dielectric tensor with the measured spectra. Hence, a numerical inversion of
the ellipsometric equations and fit to experimental data was performed for
all samples.\cite[]{AzBa77, AlTo97}

\section{Dielectric function data}\label{dielectric}

In this section, we give the dielectric tensors obtained by ellipsometry for
each compound. In general, our data are consistent with refractive index
measurements done by prism minimum deviation methods in the transparency
range of three of the compounds. \cite{BoKa71,BoKa72} We compare our results
to earlier ellipsometric measurements when available, and also with results
of normal-incidence reflectivity $R$. Our interpretation of the spectra
regarding transition energies will be given in Sect. \ref{elestr}.

\subsection{CuInSe$_2$}\label{cis}

From the four investigated compounds, CuInSe$_2$ has been the most studied
due to its applicability to photovoltaic devices.  Understanding and
modeling of solar cell performance requires a thorough knowledge of the
fundamental optical properties.  For this reason, several ellipsometric
studies of CuInSe$_2$ have already been
undertaken.\cite{KaHa83,AbHo91,BoMi97,HiLa97,KaAd98} The most complete is
the recent publication by Kawashima et al., \cite{KaAd98} where polarized
spectra from 1.2 to 5.3 eV are given. However, the important region of the
fundamental gap is outside this energy range.  Therefore, we show our
results including the gap region in Fig.~\ref{cuinsep}. We have checked
Kramers-Kronig consistency \cite{AsSt83} of these data to be better than
$\pm 0.5\%$, with larger residual structure of $\pm 2\%$ at bandgap. The
refractive indices and extinction coefficients obtained in both
polarizations are listed in Table \ref{cisnk} each 0.1 eV.  Precision and
accuracy of $\langle \varepsilon_2 \rangle$ in spectral regions of small
absorption are poor. \cite{AsSt83} Therefore, values of $k$ lower than 0.1
are considered inaccurate and are left blank.

Unpolarized measurements including the gap region were previously presented
by Hidalgo el al.,\cite{HiLa97} but their reported values of refractive
indices were somewhat low, indicative of the presence of surface overlayers.
These authors etched their samples in a Br-methanol solution, which
according to our experience does not produce the best ellipsometric spectra
for CuInSe$_2$, as already said in Sect. \ref{expt}.  References
\onlinecite{AbHo91} and \onlinecite{BoMi97} both give much too low
refractive indices, probably due to lack of attention to sample surface
quality.  On the contrary, in the work by Kazmerski et al.,\cite{KaHa83}
rather accurate values were obtained at several single wavelengths between
546 and 750 nm by correcting the ellipsometric measurements for a surface
layer of native In$_2$O$_3$ oxide.  The estimations of $n$ below bandgap of
Sobotta et al.\cite{SoNe80} are between 2.9 and 3.0 at 0.8 eV, in fair
agreement with our data. Finally, in the region above 1.2 eV, we obtain
rather similar spectra to those of Ref.~\onlinecite{KaAd98}, with only minor
differences.

The spectra of normal-incidence reflectivity $R$ in both polarizations,
calculated from $n$ and $k$ values of Table \ref{cisnk}, are plotted in
Fig.~\ref{cisref}. The labeling of the transitions has been chosen in
relationship to standard ZB notation and will be discussed in
Sect. \ref{elestr}.  In the literature, there are several
measurements of $R$ of CuInSe$_2$, either without
\cite{GaTa74,RiDe77,TuKi83} or with \cite{AuNe81} polarization dependence.
Except for the measurement by Turowski et al. \cite{TuKi83} where the values
of $R$ are quite low, the other measurements show good $R$ levels, only a
bit low towards the UV regions. The two measurements at liquid nitrogen
temperature \cite{RiDe77,AuNe81} start at 2 eV and cover a broader UV range
than our data, therefore they are quite informative in terms of observed
interband transitions. In the common energy range, we observe similar
structures and polarizations than Austinat et al. \cite{AuNe81}

\subsection{CuGaSe$_2$}\label{cgse}

Refractive index data of CuGaSe$_2$ have been reported by Boyd et al.
\cite{BoKa72} in the transparency range of the compound by prism minimum
deviation angle measurement.  Kawashima et al.  \cite{KaAd98} have
determined the dielectric tensor of CuGaSe$_2$ from 1.2 to 5.3 eV.  There is
another ellipsometric measurement by Bottomley et al.  \cite{BoMi97} that
suffers from the same shortcoming already mentioned (see Sect. \ref{cis}),
namely, an optically deficient sample surface.  Therefore, it cannot be
taken into consideration for the following discussion.

In the overlapping energy region between 1.2 and 1.6 eV, the two mentioned
sets of refractive indices (Refs. \onlinecite{BoKa72} and
\onlinecite{KaAd98}) differ by about 0.08. Also, while Boyd's \cite{BoKa72}
birefringence is considerable, it is insignificant in the measurement of
Kawashima. \cite{KaAd98} It seems possible that this discrepancy stems from
the presence of In impurities in the THM crystal used in Ref.
\onlinecite{KaAd98}. However, our measurement also of a THM crystal, shown
in Fig. \ref{cugasep}, agrees best with Boyd et al. data in both magnitudes
($n$ and birefringence). Our spectra of $\varepsilon$ and $R$ display clear
and sharp structures, in contrast with those of Ref. \onlinecite{KaAd98},
thus a possible explanation of the difference is that the In content of our
THM sample is smaller than theirs.

Our values of $n$ and $k$ are listed in Table \ref{cgsenk}, where inaccurate
data of $k < 0.1$ have been omitted. Overall Kramers-Kronig consistency of
the dielectric functions is better than $\pm 0.3\%$ and somewhat larger
($\pm 1\%$) around bandgap. Figure \ref{cgseref} displays the polarized
reflectivities calculated from our data. The transitions are labeled
according to the assignments done in Sect. \ref{elestr}. There are two
published measurements of $R$ of this compound which show lower values. The
unpolarized spectrum at room temperature of Turowski et al. \cite{TuKi83} up
to 8 eV shows many structures but it gives too low $R$ values and is quite
deformed above 4 eV. The polarized spectra measured at low temperature by
Matveev et al. \cite{MaGr96} are somewhat better but are restricted to the
1.7 to 4.5 eV energy range.

\subsection{CuInS$_2$}\label{cins}

The optical properties of CuInS$_2$ in the range of transparency of the
compound have been investigated by Boyd et al. \cite{BoKa71} In the opaque
region, polarized reflectivity spectra around the gap have been reported by
Makarova et al. \cite{MaMe88} and in a wider energy range by Syrbu et al.,
\cite{SyCr98} both at room and liquid nitrogen temperatures.

Figure \ref{cinsep} shows the result of our measurements together with data
taken from Ref. \onlinecite{BoKa71}. There is no overlap between both sets
of data but the endpoints just coincide. As it is seen in Fig. \ref{cinsep},
the values of $\langle\varepsilon_1\rangle$ and therefore of refractive
index do not join smoothly. Our values of $n$ listed in Table \ref{cinsnk}
are about 0.05 higher and our birefringence $\Delta n = n_{\parallel} -
n_{\perp}$ is slightly smaller. Also, the absorption edge in our crystal is
located at lower energy. Although the origin of these differences is not
clear, they may be caused by a variation of stoichiometry. \cite{Hs86}
Kramers-Kronig consistency of the dielectric functions in this case is quite
good ($\pm 0.1\%$), increasing to about $1\%$ at the edges of the spectra.

Our reflectivity, given in Fig. \ref{cinsref}, is higher than
those reported earlier. This is due to careful surface preparation. In Ref.
\onlinecite{MaMe88} the authors measured $n = 2.55$ and $k=0.59$ by
ellipsometry at 1.96 eV (He-Ne laser). These values are clearly indicative
of an optically not abrupt surface, in spite that the sample was freshly
polished and etched in CCl$_4$ prior to the measurement. Therefore, this
treatment does not seem quite adequate. The $R$ spectra of Syrbu et al.
\cite{SyCr98} show somewhat low values. Nevertheless, these spectra contain
fine structure even at room temperature. The gross features resemble those
seen in our spectra.

\subsection{CuGaS$_2$}

 The optical properties of CuGaS$_2$ have been reviewed recently by Rife.
\cite{Ri98} Refractive indices in the transparency range were
measured by Boyd et al. \cite{BoKa71} At higher energies, from 2.5 to 26 eV,
the optical functions \cite{Ri98} were calculated from Kramers-Kronig
analysis of reflectivity data measured at 80 K. \cite{RiDe77} Comparing our
results with those available data we find very good agreement with Boyd et
al. \cite{BoKa71} (see Fig. \ref{cgasep}). Our values of $n$, given in Table
\ref{cgasnk}, are slightly higher and the birefringence smaller due to the
small In content of our crystal. On the contrary, the optical functions in
the opaque range given by Rife \cite{Ri98} are substantially different from
our data. In the overlap region our values of $n$ are a 15\% higher in
average, discounting excitonic peaks. Also, our value of $k$ above gap (f.
i., at 3 eV) is approximately a factor of 2 smaller than given in
Ref.~\onlinecite{Ri98}. These differences can be caused by the original
values of $R$ used \cite{RiDe77} that are 8 to 10\% lower (excepting
prominent structures) relative to the $R$ calculated from our data and shown
in Fig.~\ref{cgasr}. The different sample temperature of the measurements is
not likely to produce these differences. For instance, dispersion of
birefringence of CuGaS$_2$ near the absorption edge does not vary much
between room and liquid He temperatures, \cite{ChEo97} the largest
difference in behavior being given by the shift of the band gap.

The spectra shown in Fig.~\ref{cgasep} are consistent under Kramers-Kronig
transformations within $\pm 0.5\%$, with larger residual structures of $\pm
1\%$ at the band gaps. Concerning the structures observed in the spectra,
the $R$ spectra of Rife et al. \cite{RiDe77} at 80 K comprise a wide energy
range, thus giving important information about electronic transition
energies.

\section{Critical point analysis}\label{cps}

Different aspects of the optical and electronic properties of Cu-III-VI$_2$ compounds
have been investigated by several groups. Here we concentrate on the
optical transition energies and their relationship to the electronic band
structure. The band structure calculations of Jaffe and Zunger
\cite{JaZu83,JaZu84} provide a theoretical reference framework for these
class of compounds. However, from the experimental point of view, such a
comprehensive framework is missing. The structure of the fundamental
absorption edge is quite well understood but no unanimous interpretation of the upper
transitions has yet been established. In part, this is due to the fact that
many experimental works were done before the mentioned {\em ab-initio}
calculations could be realized. But, in addition, studies done after
those calculations, have seldom attempted to assign the optical transitions
above the fundamental edge. Also, some of the assignments were done without
taking into account the selection rules of the transitions. In this work,
we admit the complexity of such assignments, but we look for general trends
in the spectra and give a consistent view that agrees with symmetry
arguments.

The dielectric function of a semiconductor is closely linked to its
electronic band structure. The features observed in $\varepsilon(\omega)$ at
optical energies are related to interband transitions characterized by large
or singular joint density of states (DOS), i.~e., critical points (CPs). The
behavior of  $\varepsilon(\omega)$ near a CP is given by \cite{Ca69,LaGa87}
\begin{equation}\label{linsh}
\varepsilon(\omega) = C- A e^{i\phi} (\omega - E + i\gamma )^n,
    \end{equation}
where $A$ is the amplitude, $\phi$ the phase angle, $E$ the energy
threshold, and $\gamma$ the broadening. The exponent $n$ takes the values
$-\frac{1}{2}$, 0, and $\frac{1}{2}$ for one (1D), two (2D), and
three-dimensional (3D) CPs, respectively. Discrete excitons (0D) are
represented by $n=-1$. Conclusions about the bands can be drawn by
evaluating experimental $\langle\varepsilon(\omega)\rangle$ spectra using
Eq. \ref{linsh} to determine CP parameters. Usually, fitting procedures are
run on numerically calculated derivatives of
$\langle\varepsilon(\omega)\rangle$. Here, we have calculated the $d^2
\langle\varepsilon\rangle /d\omega^2$ of our experimental tensor components
using the standard technique of smoothing polynomials. \cite{LaGa87}
Appropriate polynomial degree and number of correlated points were chosen to
avoid line shape distortion while giving the best possible structure
enhancement. For the fundamental band gap features the best fits were
obtained with excitonic line shapes for all three transitions. For the other
strong structures, 2D line shapes were suitable. Then, for weaker
structures, 2D line shapes were used as well. The obtained derivatives along
with their best fits are presented in Figs. \ref{cisdd} to
\ref{cugdd}.

As it happens with the spectra of  $\langle\varepsilon(\omega)\rangle$, the
second-derivative spectra of both selenides (Figs. \ref{cisdd} and
\ref{cgsdd}) bear close resemblance to each other, as do both sulfides (Figs.
\ref{cuidd} and \ref{cugdd}).
At room temperature, the former spectra display more prominent structures
than the latter. In general, spectra of ordinary polarization ({\bf E}
$\perp c$) contain more structure than the extraordinary ones ({\bf E}
$\parallel c$), except in the case of CuInS$_2$ where there is only partial
polarization selectivity and all transitions are present in both
polarizations. However, the general traits of all four spectra are alike. A
closer consideration of the electronic structure of these compounds is
needed in order to look for plausible assignments for the observed
transitions.

\section{Electronic structure: Assignments and discussion}\label{elestr}

\subsection{Particularities of the electronic structure}

Ternary chalcopyrites I-III-VI$_2$ can be viewed as isolectronic analogs of
the II-VI binary semiconductors. The symmetry reduction given by the
chemical difference between the two cations, combined with the two
structural modifications $\eta$ (tetragonal distortion of the unit cell) and
$u$ (anion displacement from the ideal tetragonal site) result in a richer
range of physical and chemical properties than their binary analogs.
Intricacy is further enhanced in Cu-III-VI$_2$ compounds where noble-atom
$d$ orbitals strongly participate in bonding through hybridization with the
anion $sp$ states.

In the simplest approach where only symmetry differences are considered, the
electronic structure of CH can be derived from that of ZB binary analogs.
\cite{CaAr90}  The Bravais lattice of CH is shown in Fig. \ref{cell}. The
corresponding elementary cell contains eight atoms
(Cu$_2$-III$_2$-VI$_4$) instead of the two found in the binaries.
Consequently the Brillouin zone  reduces its volume by a factor of four. Sets
of four different wavevectors of the original ZB Brillouin zone fold into a
single point of the new, four times smaller, CH Brillouin zone. Both
Brillouin zones are depicted in Fig. \ref{BZ}. The main
symmetry points of the CH Brillouin zone are (in units of $\pi/a$):
$\Gamma(000)$ with states originated in $\Gamma(000)$, $X(002)$, $W(201)$,
and $W(021)$; $T(001)$ with states from $\Delta(001)$, $\Delta(00\bar{1})$,
$X(200)$, and $X(020)$; and $N(110)$ with states from $L(111)$,
$L(\bar{1}\bar{1}1)$, $\Sigma(1\bar{1}0)$, and $\Sigma(\bar{1}10)$. This
change in symmetry also forces degeneracy of some electronic states, either
directly ($N$ states are always doubly degenerate) or relating spatially
uncoupled electronic states by means of time reversal symmetry [as for
$(T_1+T_2)$ and $(T_3+T_4)$]. At the same time some existing degeneracies of
the ZB electronic states are apt to be lifted.

The relevance of these symmetry facts depends on the actual value of the
tetragonal interaction. The crystal field breaks the degeneracy of the
topmost valence band states and induces the splitting of the ZB $\Delta$
states at the new $T$ Brillouin zone edge states. The details of tetragonal
distortion effects on the symmetry of electronic states with energies close
to the fundamental band gap are given in Fig. \ref{folding}. For
convenience, in the remaining part of this work, we shall term B[A] for the
link between $k$-points A and B, in ZB and CH compounds, respectively.

The tetragonal perturbation also changes the interaction between atomic
states that compose the valence and conduction bands. In a wide-gap II-VI
semiconductor the valence band is mainly built from $s$ and $p$ states of
the VI-anion. The $s$ states form a band at about 11 eV below the topmost
valence band \cite{StEu69a} and are therefore irrelevant for the
experimental energy range considered in this work. The $p$ states span a
range of about 5 eV. In the binary analogs these states have $\Gamma_{15}$
symmetry or, if lower, compatible with it. For example, at the center of the
Brillouin zone, in a CH structure, the valence band states of a II-VI
compound have $\Gamma_{15v} + (X_{5v}+X_{3v}) + 2(W_{3v} + W_{2v} + W_{1v})$
symmetry, equivalent to having 4 $\Gamma_{15}$. In the ternary Cu-III-VI$_2$
compounds, Cu-$3d$ states reside in valence band energy range. The $d$
states split into two $\Gamma_{12}$ and three $\Gamma_{15}$ states in the
tetrahedral ZB symmetry. Only $\Gamma_{15}(d)$ states can interact with the
anion $p$ states giving rise to bonding and antibonding bands, whereas the
$\Gamma_{12}(d)$ states form the nonbonding band. The associated DOS of
these three bands has been observed in photoemission experiments.
\cite{TuMa85,TaKa92} If we now reduce the symmetry to that of CH we get new
coupling possibilities. At the Brillouin zone center the 12 VI-anion $p$
states (three for each of the four atoms in the elementary cell) reduce to
($\Gamma_{4v}+\Gamma_{5v}$)[$\Gamma_{15v}$] + $\Gamma_{5v}$[$X_{5v}$] +
$\Gamma_{2v}$[$X_{3v}$] + $\Gamma_{5v}$[$2W_{3v}$] +
($\Gamma_{3v}+\Gamma_{4v}$)[$2W_{2v}$] +
($\Gamma_{1v}+\Gamma_{2v}$)[$2W_{1v}$], that is
$\Gamma_{1v}+2\Gamma_{2v}+\Gamma_{3v}+\Gamma_{4v}+3\Gamma_{5v}$. The six
$\Gamma_{15}(d)$ Cu states split into ($\Gamma_{3}+\Gamma_{4}+2\Gamma_{5}$)
and the four $\Gamma_{12}(d)$ states into
($\Gamma_{1}+\Gamma_{2}+\Gamma_{3}+\Gamma_{4}$). Consequently a coupling
between anion $p$ states and $\Gamma_{12}(d)$ states is also possible. If
these states have energies that lie near the middle of the valence band, the
most sensitive to hybridization would be $\Gamma_{2}$ and $\Gamma_{3}$
states, because they are closer to $p$ states with alike symmetry. The
inclusion of Cu-3$d$ states does not change significantly the generic
diagram of energy levels displayed in Fig.~\ref{folding}. It merely adds a
new fourfold band with small dispersion corresponding to $\Gamma_{12}(d)$ Cu
states.

This symmetry predicted scheme for the electronic valence band structure of
I-III-VI$_2$ CHs is confirmed by photoemission spectroscopy
\cite{TuMa85,TaKa92} and reproduced by theoretical calculations \cite{JaZu83}. Both
results conclude that the upper valence band is made exclusively by $p-d$
hybridization of Cu and VI group anions, whereas the III group cations do
not contribute. Structures in the $d$-DOS are almost insensitive to
substitutions of the III group cation. The strength of
$\Gamma_{15}(d)$--$\Gamma_{15}(p)$ interaction depends inversely on the
energy separation between Cu-3$d$ orbitals and VI-anion $p$ orbitals. This
repulsive interaction pushes the antibonding $p-d$ states to higher
energies and the resulting valence band width is narrower for heavier VI-atoms.
Moreover, not all these antibonding $p-d$ states are consumed in the
valence band and a significant amount of anion $p$ character exists also at
the conduction band. This is obviously accompanied by some hybridized Cu-$d$
character. Tetragonal crystal field leaves the $\Gamma_{12}(d)$ Cu states
in a narrow, almost unhybridized band midway of the bonding
and antibonding $p-d$ bands.

For a complete description of the system, the atomic spin-orbit interaction
should be added. The symmetry analysis would change and the coupling between
atomic electronic states would differ accordingly. From the point of view of
symmetry, the spin-orbit interaction leads to further level splittings and
to less selective polarization dependence of the transitions. However,
experimentally, the only manifestation of spin-orbit interaction is the well
known fundamental gap triplet (see below) clearly seen in CuGaSe$_2$. The
$p-d$ hybridization is known \cite{ShWe75} to reduce the spin-orbit effects
relative to the ZB analogs, so that, in sulfides, the effective spin-orbit
parameter is very small (see Table \ref{fgap}). Because the general traits
of the higher transitions look similar for all four compounds studied in
this work, we believe that the spin-orbit interaction is not meaningful
above the band gap. Therefore, the complexity introduced by the spin-orbit
interaction has been omitted in our subsequent analysis of the optical
properties of Cu-III-VI$_2$ CHs.

\subsection{Properties of the optical transitions}

In a first approximation, the optical functions of ternary compounds (see
Figs. \ref{cuinsep}, \ref{cugasep}, \ref{cinsep}, and \ref{cgasep}) are
similar to those of the binary analogs. Nevertheless, symmetry differences
between ZB and CH structures and the contribution of Cu-3$d$ states to the
upper valence band do result in distinctive features in the optical spectra.
The main significative traits are described in the following.

The structure of the fundamental absorption edge of these compounds is well
known. \cite{ShWe75} The crystal field interaction splits the threefold
degenerate $\Gamma_{15}$ valence band maximum, as shown in Fig.
\ref{folding}. Considering besides the spin-orbit interaction, the
fundamental gap consists of three transitions $E_0(A) \equiv E_A$, $E_0(B)
\equiv E_B$, and $E_0(C) \equiv E_C$. From symmetry arguments only
transition $E_0(B)$ is forbidden in {\bf E} $\parallel c$ polarization.
However, $E_0(A)$ and $E_0(C)$ transitions are mainly seen in {\bf
E} $\parallel c$ and {\bf E} $\perp c$, respectively. The energies and
selection rules found from experiment allow to calculate the energetic
disposition of the three valence band states and the crystal field
($\Delta_{cf}$) and spin-orbit ($\Delta_{so}$) parameters using the
quasicubic model.  \cite{ShWe75} Compared with the binary analogs,
Cu-III-VI$_2$ ternaries show a significant band gap reduction due to
repulsive interaction between Cu-$3d$ states and VI-anion $p$ states.
\cite{JaZu84}

Above the fundamental gap, the dielectric function in the binaries is mainly
dominated by two strong transitions, $E_1$ and $E_2$, and a third less
active response $E'_0$. \cite{AdTa91,JaGu94} In our description of the
interband transitions we follow the standard notation where the numeric
subindex describes the Brillouin zone region where the transition
originates. In Fig.~\ref{folding} we show the ZB states involved in those
transitions. The $E'_0$ structure corresponds to the
$\Gamma_{15v}\to\Gamma_{15c}$ transition which in II-VI compounds is usually
found above $E_1$ and $E_2$ and occurs beyond our experimental range. In the
CH structure the Brillouin zone gathers different $k$-points of a folded ZB
Brillouin zone and reduced symmetry can induce electronic transitions that
were weak or forbidden in the binaries. Examples are indirect transitions
like $\Gamma(000)\to X(002)$, or the enhaced joint DOS at the $T(001)$ point
coming from ZB $\Delta(001)\to\Delta(001)$. Thus, there is an increase in
the number of symmetry allowed interband transitions and consequently the
optical spectra of ternary compounds are richer in structure.  Several,
usually weak, transitions are expected, superimposed on the dominant
spectral features (see Figs. \ref{cisdd}--\ref{cugdd}) stemming from $E_1$
and $E_2$ transitions of the binary compounds. Another important effect of
tetragonal symmetry is polarization selectivity which proves very helpful to
assign observed transitions. For instance, transitions at the $N$ point are
allowed in both polarizations, whereas transitions involving former ZB
$X$-point states show a strong anisotropy at the $\Gamma$ and $T$ points of
CH Brillouin zone. The selection rules for dipolar electric transitions at
high symmetry points of CH Brillouin zone are summarized in
Table~\ref{selrules}.

The contribution of Cu-$3d$ states to the upper valence band affects only
slightly the optical spectra, its main contribution being the band gap
reduction and suppression of spin-orbit effects. Transitions from the
nonbonding $\Gamma_{12}(d)$ Cu states to the conduction band are forbidden
in the ZB structure, but allowed for the CH structure. Nevertheless, if
present, these transitions should be very weak because theoretical
calculations \cite{JaZu83} and photoemission experiments
\cite{TaKa92,TuMa85} show that $\Gamma_{12}$ states form a very narrow band
with a small dispersion induced by the tetragonal interaction.

\subsection{Assignment of optical transitions}

In the well-known region of the fundamental gap, our measured transition
energies are gathered in Table \ref{fgap} along with relevant data published
before. Literature results at low temperatures are given in the cases where
small splittings cannot be resolved at room temperature. The energies found
in this work for both CuInVI$_2$ compounds compare well with the reference
values. In both cases we find $E_0(A)=E_0(B)$ within experimental error, as
is habitual at room temperature. The gap of 1.04 eV for CuInSe$_2$ at room
temperature is rather high, indicating a proper stoichiometry of the
crystal. \cite{ShCh97a} The gap we measure for CuInS$_2$ is also a good
value; the gap of the best stoichiometric CuInS$_2$ at room temperature is
considered to be 1.535 eV.\cite{Hs86}
In the two CuGaVI$_2$ compounds we find slightly reduced gaps and
$\Delta_{cf}$ parameters due to the small In content of the crystals grown
by the THM process. Comparing the measured gaps with the references we
estimate a composition CuGa$_{0.95}$In$_{0.05}$VI$_2$ for both crystals.

The transition energies above the fundamental gap obtained from ellipsometry
and low temperature polarized reflectivity measurements have been collected
in Tables \ref{enerselen} and \ref{enersulf}. By inspecting all the spectra,
we can establish a general pattern for the outstanding optical transitions
above the fundamental gap of the four Cu-III-VI$_2$ compounds analyzed in
this work. In all spectra the first strong transition, called $E_1(A)$, is
allowed in both polarizations. At $\approx 0.3$ eV above it there is a
weaker transition $E(X\Gamma)$ that appears only in perpendicular
polarization. About 0.8 eV above $E_1(A)$ there is another optical
transition allowed in both polarizations, labeled $E_1(B)$. Nearby, and only
in parallel polarization, emerges a transition $E(\Delta X)$, located at
$\approx 0.5$ eV above (VI = Se) or below (CuGaS$_2$) $E_1(B)$. Close to 5
eV a strong double structure is observed: $E_2(A)$ and $E_2(B)$ allowed in
perpendicular and parallel polarization, respectively. This general pattern
is also in agreement with ellipsometric measurements reported for
CuAlSe$_2$.\cite{AlPa00}

We associate transitions $E_1(A)$ and $E_1(B)$ to $E_1$-like transitions at
the $N$ point of the Brillouin zone. As depicted in Fig. \ref{folding}, the
$E_1$ transition of binary ZB splits into two $N_{1v} \to N_{1c}$
transitions in CH. If we identify $E_1(A)$ and $E_1(B)$ with this
pair, the splitting between the two $N_{1v}$ involved valence band states
would be of the order of 0.8 eV. Due to the proximity of another band coming
from $\Sigma$ points in ZB, theory \cite{JaZu83} gives three close
$N_{1v}$ valence band states. Calculated energy differences are of the order
of $\Delta E (N_{1v}^{(1)}-N_{1v}^{(2)})
\approx$ 0.4 eV and $\Delta E (N_{1v}^{(2)}-N_{1v}^{3)}) \approx$ 0.6 eV
(except in CuGaS$_2$ where they are 0.2 eV and 0.8 eV, respectively).
All three possibilities give the correct order of magnitude of the measured
0.8 eV. However, we prefer the assignment of transitions $E_1(A)$ and
$E_1(B)$ to the lowest-energy
$N_{1v}^{(1)}\to N_{1c}^{(1)}$ and $N_{1v}^{(2)}\to N_{1c}^{(1)}$, respectively.

$E(X\Gamma)$ is a new interband transition, only allowed in {\bf E}$\perp
c$, with no corresponding direct transition in binary compounds. Using the
diagram
of Fig. \ref{folding} the three possible assignments by symmetry
are the pseudodirect transitions $E(\Gamma X)$:
$\Gamma_{5v}^{(1)}[\Gamma_{15v}]\to\Gamma_{3c}[X_{1c}]$, $E(X\Gamma)$:
$\Gamma_{5v}^{(2)}[X_{5v}]\to\Gamma_{1c}[\Gamma_{1c}]$, and $E'(\Gamma
X)$: $\Gamma_{5v}^{(1)}[\Gamma_{15v}]\to\Gamma_{2c}[X_{3c}]$. Calculations
\cite{JaZu83} predict for most of the four Cu-III-VI$_2$ compounds energies
in the sequence $E(\Gamma X) < E_1(A) < E(X\Gamma) < E'(\Gamma X)$.
Following this theoretical prediction, we propose to assign
$\Gamma_{5v}^{(2)}\to\Gamma_{1c}$ to the $E(X\Gamma)$ optical structure.
This feature on the high energy side of $E_1(A)$ corresponds to an interband
transition between the heavy hole $p$ band and the bottom of the conduction
band. Notice also that in all experimental spectra we find a weak shoulder
below $E_1(A)$ which is only allowed in {\bf E} $\perp c$ . We propose to
associate this shoulder to the mentioned lower energy $E(\Gamma
X)$ optical transition. Also, in the two CuIn-VI$_2$ compounds there is
another transition only allowed in {\bf E} $\perp c$ that we assign to  $E'(\Gamma X)$.

The structure that appears in {\bf E} $\parallel c$ and is labeled $E(\Delta
X)$ has no correspondent direct transition in the binary analogues. Taking
into account both selection rules and calculated energies, \cite{JaZu83} the only
matching transition from the upper valence band to the conduction band would
be the pseudodirect transition $(T_{3v}+T_{4v})[\Delta_{3v}+\Delta_{4v}] \to
(T_{1c}+T_{2c})[X_{1c}]$. Yet another possibility could be to associate this
structure to electronic transitions from nonbonding $\Gamma_{12}(d)$ states
to the minimum of the conduction band at $\Gamma_{1c}$. However, if we use
the experimental values of measured maximum DOS of nonbonding $\Gamma_{12}(d)$
states \cite{TuMa85,TaKa92} to calculate the expected energies of such a
transition, we obtain energies that do not coincide with our
experimental $E(\Delta X)$, even if we consider broadening effects on $\Gamma_{12}(d)$
states. Also, comparing CuInSe$_2$ and CuGaSe$_2$ where this transition is
particularly well resolved, the difference between
both $E(\Delta X)$ energies
should coincide with the difference in band gaps, \cite{TuMa85} which is not
the case.
Then, we discard that  unhybridized  $\Gamma_{12}(d)$ states are involved in this
transition and conclude that within the spectral range covered by our
experimental set-up, only $p-d$ hybridized anti-bonding valence band states
contribute to the main band-to-band electronic transitions.

The four compounds show a high dielectric response and large anisotropy at
$\approx 5$ eV. In analogy with II-VI compounds, we identify the observed
structures $E_2(A)$ and $E_2(B)$ with $E_2$ transitions. Within the energy
range of $E_2$ transitions, notice that the $X(002)$ point folds to
the $\Gamma$ point, and the other two equivalent points in ZB,
$X(200)$ and $X(020)$, fold to the $T$ point. The $X$ direct
transition at $\Gamma$, $\Gamma_{5v}^{(2)}[X_{5v}]\to\Gamma_{3c}[X_{1c}]$,
is only allowed in perpendicular polarization. On the contrary,
$X(200)$ and $X(020)$ states are coupled at $T$-point. The new
electronic states give rise to a pair of direct transitions, $E_2(A)$:
$(T_{3v}+T_{4v})[X_{5v}]\to T_{5c}^{(1)}[X_{1c}]$, and $E_2(B)$:
$T_{5v}[2X_{5v}]\to T_{5c}^{(1)}[X_{1c}]$, allowed in perpendicular and
parallel polarization, respectively. Theory predicts for the two valence
band states at $T$, $T_{5v}$ and $(T_{3v}+T_{4v})$, a splitting of about
1--1.5 eV. $T_{5v}$ belongs to the upper antibonding manifold bands while
$(T_{3v}+T_{4v})$ belongs to the $p-d$ bonding energy region. According to
theoretical predictions, only the transitions $T_{5v}\to T_{5c}^{(1)}$ would
contribute to $E_2$ (the energy of transition
$\Gamma_{5v}^{(2)}\to\Gamma_{3c}$ is always above that of $T_{5v}\to
T_{5c}^{(1)}$). This seems to be in contradiction with experimental results,
which shows that transitions with {\bf E}$\parallel c$ are also allowed in
this energy region. The discrepancy should be overcome if the splitting of
the two valence bands is of $\approx 0.2$ eV, much smaller than
calculated.\cite{JaZu83} But notice also that at $T$ point, the energy
difference between $T_{5c}^{(1)}$, and $(T_{1c}+T_{2c})$, is only of $\approx
0.3$ eV (except for CuGaSe$_2$, which is $\approx 0.03$ eV), and the doublet
$T_{5v}\to (T_{1c}+T_{2c})$ (allowed in {\bf E}$\perp c$) and
$T_{5v}\to T_{5c}^{(1)}$ (allowed in {\bf E}$\parallel c$), can be also a good
candidate for $E_2(A)$ and $E_2(B)$ transitions.
The proposed assignments and notation of the main optical transitions are given in
the generic band structure displayed in Fig. \ref{bandes}. Although we
cannot distinguish the origin of the observed features in $k$-space, the
main contributions are drawn at zone center $\Gamma$ and zone edge N and T
points.

\section{Summary}\label{summary}

We have presented the dielectric tensor components of the four ternary
chalcopyrites CuInSe$_2$, CuGaSe$_2$, CuInS$_2$, and CuGaS$_2$, measured on
single crystal samples at room temperature in the energy range from 1.4 to
5.2 eV (from 0.9 eV for CuInSe$_2$). The  pseudodielectric components
have been obtained from complex reflectance ratios measured in appropriate
configurations. We have paid special attention to the problem of preparing
and maintaining a good sample surface throughout the experiments. Thus, the
obtained dielectric function values are representative of the bulk material.
This is confirmed by the excellent agreement of our results with those of
earlier prism minimum deviation methods in the transparency range of three
of the compounds.

In addition, we have obtained the parameters of interband transitions from
the numerically differentiated components. In particular, we have identified
general trends of the spectra and given assignments for the most
important transitions, taking into account band structure calculations and
the appropriate selection rules for coupling between electronic states. Within the
spectral range covered by our experimental set-up, only $p-d$ hybridized
anti-bonding valence band states contribute to the main band-to-band
electronic transitions. Hence, the optical spectra of these compounds
ressemble more than previously assumed those of their ZB analogues.

Both the spectral dependence of the optical functions and the critical point
analysis are expected to be useful in further studies of structures based on
these compounds.

\acknowledgments
This work has been partially supported by the Spanish CICYT project
TIC97--0594.


\newpage

\begin{table}[htbp!]
\caption{Values of refractive indices $n$ and extinction coefficients
$k$ of CuInSe$_2$ at intervals of
0.1~eV.
}
\label{cisnk}
\begin{tabular*}{0.4\textwidth}{@{\extracolsep{\fill}}*{5}{ccccc}}
\toprule
E(eV) & n$_\perp$ & k$_\perp$     & n$_\parallel$& k$_\parallel$ \\
\colrule
0.9   &  2.937   &              &   2.950     &           \\
1.0   &  3.048   &  0.165       &   3.036     & 0.179     \\
1.1   &  3.033   &  0.314       &   3.022     & 0.320     \\
1.2   &  3.012   &  0.359       &   2.990     & 0.358     \\
1.3   &  3.003   &  0.414       &   2.982     & 0.406     \\
1.4   &  2.969   &  0.460       &   2.957     & 0.426     \\
1.5   &  2.949   &  0.479       &   2.938     & 0.452     \\
1.6   &  2.935   &  0.501       &   2.925     & 0.479     \\
1.7   &  2.931   &  0.519       &   2.920     & 0.504     \\
1.8   &  2.931   &  0.543       &   2.916     & 0.527     \\
1.9   &  2.933   &  0.571       &   2.914     & 0.550     \\
2.0   &  2.937   &  0.604       &   2.922     & 0.573     \\
2.1   &  2.941   &  0.637       &   2.936     & 0.593     \\
2.2   &  2.949   &  0.671       &   2.953     & 0.625     \\
2.3   &  2.960   &  0.712       &   2.971     & 0.665     \\
2.4   &  2.974   &  0.763       &   2.998     & 0.714     \\
2.5   &  2.983   &  0.828       &   3.027     & 0.773     \\
2.6   &  2.993   &  0.908       &   3.072     & 0.853     \\
2.7   &  2.988   &  1.003       &   3.125     & 0.983     \\
2.8   &  2.951   &  1.119       &   3.095     & 1.223    \\
2.9   &  2.848   &  1.225       &   2.867     & 1.390    \\
3.0   &  2.709   &  1.264       &   2.635     & 1.378    \\
3.1   &  2.620   &  1.251       &   2.500     & 1.271    \\
3.2   &  2.541   &  1.236       &   2.464     & 1.169    \\
3.3   &  2.488   &  1.185       &   2.475     & 1.108    \\
3.4   &  2.479   &  1.158       &   2.505     & 1.092    \\
3.5   &  2.479   &  1.164       &   2.531     & 1.111    \\
3.6   &  2.457   &  1.200       &   2.531     & 1.161    \\
3.7   &  2.390   &  1.199       &   2.482     & 1.181    \\
3.8   &  2.355   &  1.159       &   2.471     & 1.154    \\
3.9   &  2.346   &  1.120       &   2.495     & 1.164    \\
4.0   &  2.366   &  1.081       &   2.516     & 1.207    \\
4.1   &  2.411   &  1.061       &   2.517     & 1.269    \\
4.2   &  2.473   &  1.069       &   2.482     & 1.321    \\
4.3   &  2.536   &  1.119       &   2.450     & 1.352    \\
4.4   &  2.586   &  1.194       &   2.433     & 1.380    \\
4.5   &  2.617   &  1.296       &   2.410     & 1.423    \\
4.6   &  2.613   &  1.427       &   2.391     & 1.468    \\
4.7   &  2.545   &  1.562       &   2.349     & 1.526    \\
4.8   &  2.429   &  1.649       &   2.299     & 1.583    \\
4.9   &  2.319   &  1.674       &   2.224     & 1.628    \\
5.0   &  2.251   &  1.672       &   2.145     & 1.646    \\
5.1   &  2.213   &  1.699       &   2.092     & 1.656    \\
5.2   &  2.154   &  1.750       &   2.042     & 1.688    \\
\botrule
\end{tabular*}
\end{table}

\begin{table}[htbp!]
\caption{Values of $n$ and $k$ of CuGaSe$_2$ at intervals of 0.1~eV.
}
\label{cgsenk}
\begin{tabular*}{0.4\textwidth}{@{\extracolsep{\fill}}*{5}{ccccc}}
\toprule
E(eV) & n$_\perp$ & k$_\perp$     & n$_\parallel$& k$_\parallel$ \\
\colrule
1.4   &   2.904   &               &   2.920      &            \\
1.5   &   2.942   &               &   2.968      &            \\
1.6   &   3.000   &               &   3.054      &            \\
1.7   &   3.082   &               &   3.067      & 0.200      \\
1.8   &   3.080   & 0.184         &   3.048      & 0.245      \\
1.9   &   3.102   & 0.228         &   3.065      & 0.276      \\
2.0   &   3.102   & 0.294         &   3.068      & 0.311      \\
2.1   &   3.104   & 0.331         &   3.076      & 0.338      \\
2.2   &   3.116   & 0.365         &   3.093      & 0.364      \\
2.3   &   3.137   & 0.392         &   3.114      & 0.393      \\
2.4   &   3.160   & 0.432         &   3.139      & 0.430      \\
2.5   &   3.188   & 0.476         &   3.163      & 0.472      \\
2.6   &   3.230   & 0.522         &   3.204      & 0.515      \\
2.7   &   3.267   & 0.589         &   3.251      & 0.564      \\
2.8   &   3.306   & 0.673         &   3.300      & 0.640      \\
2.9   &   3.335   & 0.784         &   3.342      & 0.749      \\
3.0   &   3.343   & 0.929         &   3.387      & 0.908      \\
3.1   &   3.318   & 1.092         &   3.363      & 1.162      \\
3.2   &   3.202   & 1.246         &   3.116      & 1.336      \\
3.3   &   3.024   & 1.327         &   2.876      & 1.308      \\
3.4   &   2.879   & 1.319         &   2.757      & 1.193      \\
3.5   &   2.778   & 1.297         &   2.748      & 1.094      \\
3.6   &   2.701   & 1.231         &   2.784      & 1.043      \\
3.7   &   2.700   & 1.176         &   2.844      & 1.036      \\
3.8   &   2.716   & 1.161         &   2.903      & 1.074      \\
3.9   &   2.734   & 1.177         &   2.945      & 1.156      \\
4.0   &   2.733   & 1.223         &   2.941      & 1.277      \\
4.1   &   2.687   & 1.253         &   2.850      & 1.364      \\
4.2   &   2.660   & 1.250         &   2.796      & 1.378      \\
4.3   &   2.650   & 1.246         &   2.765      & 1.412      \\
4.4   &   2.659   & 1.242         &   2.731      & 1.453      \\
4.5   &   2.687   & 1.255         &   2.689      & 1.497      \\
4.6   &   2.725   & 1.294         &   2.634      & 1.529      \\
4.7   &   2.764   & 1.372         &   2.585      & 1.542      \\
4.8   &   2.760   & 1.487         &   2.546      & 1.558      \\
4.9   &   2.705   & 1.608         &   2.512      & 1.582      \\
5.0   &   2.601   & 1.706         &   2.481      & 1.612      \\
5.1   &   2.476   & 1.748         &   2.443      & 1.653      \\
5.2   &   2.343   & 1.731         &   2.368      & 1.698      \\
\botrule
\end{tabular*}
\end{table}

\begin{table}[htbp!]
\caption{Values of $n$ and $k$ of CuInS$_2$ at intervals of 0.1~eV.
}
\label{cinsnk}
\begin{tabular*}{0.4\textwidth}{@{\extracolsep{\fill}}*{5}{ccccc}}
\toprule
E(eV) & n$_\perp$ & k$_\perp$     & n$_\parallel$& k$_\parallel$ \\
\colrule
1.4   &   2.874   &    0.219      &   2.866      &   0.199    \\
1.5   &   2.945   &    0.352      &   2.927      &   0.341    \\
1.6   &   2.796   &    0.422      &   2.784      &   0.405    \\
1.7   &   2.761   &    0.415      &   2.748      &   0.400    \\
1.8   &   2.742   &    0.419      &   2.727      &   0.408    \\
1.9   &   2.725   &    0.437      &   2.711      &   0.418    \\
2.0   &   2.717   &    0.449      &   2.705      &   0.426    \\
2.1   &   2.708   &    0.455      &   2.702      &   0.431    \\
2.2   &   2.708   &    0.469      &   2.708      &   0.441    \\
2.3   &   2.714   &    0.480      &   2.715      &   0.454    \\
2.4   &   2.721   &    0.499      &   2.726      &   0.471    \\
2.5   &   2.734   &    0.523      &   2.743      &   0.493    \\
2.6   &   2.747   &    0.557      &   2.767      &   0.522    \\
2.7   &   2.764   &    0.587      &   2.789      &   0.555    \\
2.8   &   2.779   &    0.635      &   2.809      &   0.602    \\
2.9   &   2.782   &    0.686      &   2.821      &   0.662    \\
3.0   &   2.783   &    0.744      &   2.828      &   0.733    \\
3.1   &   2.774   &    0.807      &   2.816      &   0.818    \\
3.2   &   2.738   &    0.870      &   2.767      &   0.902    \\
3.3   &   2.686   &    0.914      &   2.682      &   0.955    \\
3.4   &   2.633   &    0.940      &   2.596      &   0.963    \\
3.5   &   2.589   &    0.953      &   2.545      &   0.941    \\
3.6   &   2.556   &    0.959      &   2.525      &   0.920    \\
3.7   &   2.526   &    0.959      &   2.522      &   0.908    \\
3.8   &   2.505   &    0.954      &   2.524      &   0.907    \\
3.9   &   2.493   &    0.949      &   2.534      &   0.919    \\
4.0   &   2.486   &    0.951      &   2.540      &   0.944    \\
4.1   &   2.485   &    0.949      &   2.535      &   0.972    \\
4.2   &   2.502   &    0.950      &   2.533      &   0.997    \\
4.3   &   2.521   &    0.961      &   2.529      &   1.024    \\
4.4   &   2.548   &    0.992      &   2.525      &   1.059    \\
4.5   &   2.567   &    1.037      &   2.516      &   1.095    \\
4.6   &   2.580   &    1.094      &   2.507      &   1.130    \\
4.7   &   2.581   &    1.160      &   2.499      &   1.170    \\
4.8   &   2.557   &    1.230      &   2.480      &   1.216    \\
4.9   &   2.527   &    1.303      &   2.460      &   1.268    \\
5.0   &   2.477   &    1.366      &   2.424      &   1.318    \\
5.1   &   2.410   &    1.410      &   2.381      &   1.364    \\
5.2   &   2.343   &    1.440      &   2.314      &   1.387    \\
\botrule
\end{tabular*}
\end{table}

\begin{table}[htbp!]
\caption{Values of $n$ and $k$ of CuGaS$_2$ at intervals of 0.1~eV.
}
\label{cgasnk}
\begin{tabular*}{0.4\textwidth}{@{\extracolsep{\fill}}*{5}{ccccc}}
\toprule
E(eV) & n$_\perp$ & k$_\perp$     & n$_\parallel$& k$_\parallel$ \\
\colrule
1.4   &   2.579   &               &   2.574      &            \\
1.5   &   2.590   &               &   2.588      &            \\
1.6   &   2.604   &               &   2.604      &            \\
1.7   &   2.623   &               &   2.624      &            \\
1.8   &   2.646   &               &   2.647      &            \\
1.9   &   2.675   &               &   2.677      &            \\
2.0   &   2.706   &               &   2.711      &            \\
2.1   &   2.742   &               &   2.753      &            \\
2.2   &   2.779   &               &   2.800      &            \\
2.3   &   2.822   &               &   2.858      & 0.108      \\
2.4   &   2.874   &   0.101       &   2.891      & 0.222      \\
2.5   &   2.918   &   0.207       &   2.846      & 0.262      \\
2.6   &   2.888   &   0.270       &   2.842      & 0.279      \\
2.7   &   2.884   &   0.301       &   2.847      & 0.297      \\
2.8   &   2.890   &   0.325       &   2.859      & 0.316      \\
2.9   &   2.901   &   0.348       &   2.876      & 0.334      \\
3.0   &   2.929   &   0.374       &   2.898      & 0.361      \\
3.1   &   2.955   &   0.411       &   2.926      & 0.391      \\
3.2   &   2.977   &   0.453       &   2.954      & 0.427      \\
3.3   &   3.003   &   0.505       &   2.985      & 0.474      \\
3.4   &   3.028   &   0.567       &   3.016      & 0.534      \\
3.5   &   3.048   &   0.646       &   3.045      & 0.611      \\
3.6   &   3.053   &   0.733       &   3.054      & 0.709      \\
3.7   &   3.036   &   0.823       &   3.026      & 0.813      \\
3.8   &   2.996   &   0.912       &   2.962      & 0.885      \\
3.9   &   2.935   &   0.975       &   2.900      & 0.913      \\
4.0   &   2.871   &   1.014       &   2.869      & 0.922      \\
4.1   &   2.823   &   1.033       &   2.860      & 0.938      \\
4.2   &   2.787   &   1.043       &   2.862      & 0.974      \\
4.3   &   2.769   &   1.058       &   2.855      & 1.029      \\
4.4   &   2.756   &   1.084       &   2.836      & 1.083      \\
4.5   &   2.738   &   1.120       &   2.804      & 1.140      \\
4.6   &   2.715   &   1.161       &   2.769      & 1.193      \\
4.7   &   2.682   &   1.191       &   2.714      & 1.238      \\
4.8   &   2.662   &   1.219       &   2.668      & 1.268      \\
4.9   &   2.632   &   1.264       &   2.618      & 1.295      \\
5.0   &   2.591   &   1.293       &   2.584      & 1.317      \\
5.1   &   2.550   &   1.305       &   2.541      & 1.343      \\
5.2   &   2.538   &   1.329       &   2.507      & 1.380      \\
\botrule
\end{tabular*}
\end{table}

\begin{table*}[hbtp!]
\caption{Characteristic parameters of the fundamental gap of studied
Cu-III-VI$_2$ compounds. All energies are given in eV and the numbers in
parentheses indicate the error margin of the last given decimal. Unless
otherwise indicated data are results at room temperature.}
\label{fgap}
\begin{tabular*}{\textwidth}{@{\extracolsep{\fill}}*{14}{cccccccccccccc}}
\toprule
&
\multicolumn{2}{c}{CuInSe$_2$}&&\multicolumn{3}{c}{CuGaSe$_2$}&&
\multicolumn{2}{c}{CuInS$_2$}&&\multicolumn{3}{c}{CuGaS$_2$}\\
& This work & Ref. \onlinecite{ShTe73} && This work &
Ref. \onlinecite{ShTe72} & Ref. \onlinecite{ShCh97} & &
This work & Ref. \onlinecite{ShTe72} && This work &
Ref. \onlinecite{HoYa78} & Ref. \onlinecite{HoYa78} \\
& & (77 K)& & & & && & (2 K)& & & & (20 K) \\
\colrule
$E_0(A)$ & 1.04(1) & 1.038 && 1.648(2) & 1.68 & 1.686& & 1.530(5) & 1.55 &&
2.411(2) & 2.469 & 2.497 \\
$E_0(B)$ & 1.039(3) & 1.042 && 1.717(4) & 1.76 & 1.760 && 1.530(5) & && 2.530(4)
& 2.597 & 2.625 \\
$E_0(C)$ & 1.274(6) & 1.273& & 1.920(6) & 1.96 & 1.972 && && & & & 2.635 \\
$-\Delta_{cf}$ & & -0.006& & 0.093 & 0.094 & 0.099 & & &&& 0.119& 0.128 & 0.132
\\
 $\Delta_{so}$ & 0.235 & 0.233& & 0.227 & 0.234 & 0.237& & & -0.02& & & & -0.016
\\
\botrule
\end{tabular*}
\end{table*}

\begin{table}[htbp!]
    \caption{Selection rules of the dipolar interband transitions
      at the main points of the Brillouin zone of the chalcopyrite structure.}
  \label{selrules}
\begin{tabular}{ccc}
    \toprule
BZ point   &  {\bf E} $\parallel c\quad (\Gamma_4)$ & {\bf E} $\perp c\quad (\Gamma_5)$ \\
\colrule
$\Gamma$ &  $\Gamma_1 \otimes \Gamma_4$ &   $\Gamma_1 \otimes \Gamma_5$ \\
         &  $\Gamma_2 \otimes \Gamma_3$ &   $\Gamma_2 \otimes \Gamma_5$ \\
         &  $\Gamma_5 \otimes \Gamma_5$ &   $\Gamma_3 \otimes \Gamma_5$ \\
         &                              &   $\Gamma_4 \otimes \Gamma_5$ \\
$T$      & $(T_1+T_2)\otimes(T_3+T_4)$  &   $(T_1+T_2)\otimes T_5$      \\
	 &   $T_5\otimes T_5$           &   $(T_3+T_4)\otimes T_5$      \\
         &                              &                               \\
$N$      &    $N_1\otimes N_1$          &   $N_1\otimes N_1$            \\
      \botrule
\end{tabular}
\end{table}

\begin{table}[hbtp!]
\caption{Fitted upper transition energies (in eV) and their polarization for the
two studied selenides. The
numbers in parentheses indicate the error margin of the last given decimal.
}
\label{enerselen}
\begin{tabular*}{0.4\textwidth}{@{\extracolsep{\fill}}*{5}{lccccc}}
\toprule
&\multicolumn{2}{c}{CuInSe$_2$}& & \multicolumn{2}{c}{CuGaSe$_2$} \\
\colrule
Label&{\bf E}$\parallel c$ & {\bf E}$\perp c$& &{\bf E}$\parallel c$ & {\bf E}$\perp c$ \\
\colrule
$E(\Gamma X)$ & & 2.4(1) && & 2.8(1) \\
& & 2.5 \footnotemark[1]&& & \\
$E_1(A)$ &2.821(4) & 2.901(5) && 3.127(2) & 3.247(5) \\
& 2.92 \footnotemark[1]&2.92 \footnotemark[1]&&3.28 \footnotemark[2]&3.28
\footnotemark[2]\\
& 2.92 \footnotemark[3]&2.92 \footnotemark[3]&&3.08 \footnotemark[3]&3.08
\footnotemark[3]\\
$E(X\Gamma)$ &          & 3.174(5) &&           & 3.501(4)\\
& & 3.24 \footnotemark[1] && & 3.35  \footnotemark[2] \\
$E_1(B)$&3.635(5) & 3.626(5) && 4.049(5)  & 4.03(1)  \\
& 3.72  \footnotemark[1]& 3.72 \footnotemark[1]&&4.20 \footnotemark[2]&4.16
\footnotemark[2]\\
& 3.65 \footnotemark[3] & 3.65  \footnotemark[3] && & \\
$E(\Delta X)$&4.07(5)  & & &  4.49(5)  &           \\
& 4.2 \footnotemark[1] &&&& \\
& 4.15 \footnotemark[3] &&&& \\
$E'(\Gamma X)$ & & 4.2(1) && &  \\
& & 4.4  \footnotemark[1]& & & \\
$E_2(A)$  &  & 4.71(2) &&  & 4.89(5) \\
& & 4.85  \footnotemark[1] & & & \\
& & 4.70  \footnotemark[3] & & & \\
$E_2(B)$ & 4.84(4) &  && 5.1(1) &  \\
 & 4.85  \footnotemark[1]& & &5.0 \footnotemark[3] & \\
  & 4.90  \footnotemark[3]& & & & \\
\botrule
\end{tabular*}
\footnotetext[1]{Ref. \onlinecite{AuNe81} (80 K)}
\footnotetext[2]{Ref. \onlinecite{MaGr96} (80 K)}
\footnotetext[3]{Ref. \onlinecite{KaAd98}}
\end{table}

\begin{table}[hbtp!]
\caption{Main optical transition energies (in eV) and their polarization
measured above the fundamental edge in Cu-III-S$_2$.
The numbers in parentheses indicate error margins.}
\label{enersulf}
\begin{tabular*}{0.4\textwidth}{@{\extracolsep{\fill}}*{6}{lccccc}}
\toprule
 & \multicolumn{2}{c}{CuInS$_2$}& & \multicolumn{2}{c}{CuGaS$_2$} \\
\colrule
Label & {\bf E}$\parallel c$ & {\bf E}$\perp c$& &{\bf E}$\parallel c$ & {\bf E}$\perp c$ \\
\colrule
$E(\Gamma X)$&       & 2.75(8)   &&          & 3.5(1)   \\
& 3.099 \footnotemark[1] & 3.087 \footnotemark[1]& & & \\
$E_1(A)$ & 3.27(1)   & 3.27(5)   &&   3.720(5)   & 3.85(1) \\
& 3.247 \footnotemark[1] & 3.246 \footnotemark[1] && 3.84 \footnotemark[2] &
3.28\footnotemark[2] \\
$E(X\Gamma)$ & 3.6(1)   & 3.5(1)   &&          &        \\
& 3.655\footnotemark[1] & 3.669\footnotemark[1]& && 4.20\footnotemark[2]\\
$E(\Delta X)$&   &    &&   4.15(5)   &        \\
& & && 4.40\footnotemark[2]& \\
$E_1(B)$&3.94(5)   & 3.9(1)   &&   4.63(1)   &   4.53(1) \\
& 4.053\footnotemark[1] &
4.091\footnotemark[1]&&4.70\footnotemark[2]&4.68\footnotemark[2]\\
$E'(\Gamma X)$ &4.4(1) & 4.4(2) && &  \\
$E_2(A)$&4.8(1)   & 4.7(1)   &&    &  4.91(1) \\
& 5.038 \footnotemark[1] &  && & 5.12\footnotemark[2]\\
$E_2(B)$ & 5.09(3)   & 5.05(3)   &&          &    \\
& & 5.033\footnotemark[1] && 5.14 \footnotemark[2] & \\
\botrule
\end{tabular*}
\footnotetext[1]{Ref. \onlinecite{SyCr98} (77 K)}
\footnotetext[2]{Ref. \onlinecite{RiDe77} (80 K)}
\end{table}

\newpage

\begin{figure}[htbp!]
\caption{Dielectric tensor components of CuInSe$_2$. The ordinary
({\bf E} $\perp c$) functions are plotted with solid lines, and the
extraordinary ({\bf E} $\parallel c$) with dotted lines.
Upper panel (a) shows the real parts, panel (b) the imaginary parts.}
\label{cuinsep}
\end{figure}

\begin{figure}[htbp!]
\caption{Reflectivity of CuInSe$_2$ at normal incidence calculated for the two
polarizations.}
\label{cisref}
\end{figure}

\begin{figure}[htbp!]
\caption{Ordinary (solid lines) and extraordinary (dashed lines)
dielectric tensor components of CuGaSe$_2$. Upper panel (a) shows the
real parts, panel (b) the imaginary parts. Symbols in (a) are data taken
from Ref. \protect\onlinecite{BoKa72}.%
}
\label{cugasep}
\end{figure}

\begin{figure}[htbp!]
\caption{Polarized reflectivities of CuGaSe$_2$ at normal incidence.}
\label{cgseref}
\end{figure}

\begin{figure}[htbp!]
\caption{Ordinary (solid lines) and extraordinary (dashed lines)
dielectric tensor components of CuInS$_2$. Upper panel (a) shows the
real parts, panel (b) the imaginary parts. Symbols in (a) are data taken
from Ref. \protect\onlinecite{BoKa71}.}
\label{cinsep}
\end{figure}

\begin{figure}[htbp!]
\caption{Polarized reflectivities of CuInS$_2$ at normal incidence.}
\label{cinsref}
\end{figure}

\begin{figure}[htbp!]
\caption{Ordinary (solid lines) and extraordinary (dashed lines)
dielectric tensor components of CuGaS$_2$. Upper panel (a) shows the
real parts, panel (b) the imaginary parts. Symbols in (a) are data taken
from Ref. \protect\onlinecite{BoKa71}.}
\label{cgasep}
\end{figure}

\begin{figure}[htbp!]
\caption{Polarized reflectivities of CuGaS$_2$ at normal incidence.}
\label{cgasr}
\end{figure}

\begin{figure}[htbp!]
\caption{Second-derivative spectra of CuInSe$_2$. (a) Ordinary, (b)
Extraordinary polarization. Experimental points
are plotted by symbols and their best fit is given by lines. The arrows
mark the obtained critical-point energies.
}
\label{cisdd}
\end{figure}

\begin{figure}[htbp!]
\caption{Second-derivative spectra of CuGaSe$_2$. (a) Ordinary, (b)
Extraordinary polarization. Experimental points
are plotted by symbols and their best fit is given by lines. The arrows
mark the obtained critical-point energies.
}
\label{cgsdd}
\end{figure}

\begin{figure}[htbp!]
\caption{Second-derivative spectra of CuInS$_2$. (a) Ordinary, (b)
Extraordinary polarization. Experimental points
are plotted by symbols and their best fit is given by lines. The arrows
mark the fitted critical-point energies.
}
\label{cuidd}
\end{figure}

\begin{figure}[phtb!]
\caption{Second-derivative spectra of CuGaS$_2$. (a) Ordinary, (b)
Extraordinary polarization. Experimental points
are plotted by symbols and their best fit is given by lines. The arrows
mark the fitted critical-point energies.
}
\label{cugdd}
\end{figure}

\begin{figure}[htbp!]
\caption{Crystalline chalcopyrite structure Cu-III-VI$_2$ depicted in real
space. It belongs to the space group $D_{2d}^{12}$ and is a superstructure
of zinc blende $T_d^2$.  }
\label{cell}
\end{figure}

\begin{figure}[htbp!]
\caption{Brillouin zone of chalcopyrite (CH) and its relationship to that of zinc
blende (ZB). The volume of the former is four times smaller than that of the
latter. The dotted polyhedra show the ZB reciprocal space regions that fold
into the CH Brillouin zone. Symmetry points are labelled $A_B$, where $A$
and $B$ refer to the CH and ZB symmetries, respectively.
}
\label{BZ}
\end{figure}

\begin{figure}[htbp!]
\caption{Schematic representation of energy levels and their symmetry in
zinc blende (ZB) and chalcopyrite (CH) structures.}
\label{folding}
\end{figure}

\begin{figure}[htbp!]
\caption{Proposed assignments and notations for the transitions observed in
Cu-III-VI$_2$ chalcopyrites in the optical range, depicted on a generic band
structure. Dashed and solid arrows represent optical transitions allowed in
{\bf E}$\parallel c$ and {\bf E}$\perp c$, respectively. Only one of the
possible origins of the observed E$_2$-type transitions is indicated. }
\label{bandes}
\end{figure}

\end{document}